\documentclass[11pt]{article}

\usepackage{graphicx}
\usepackage{epstopdf}
\usepackage{amsmath}
\usepackage{amssymb}
\usepackage{amsfonts}
\usepackage{amsthm}
\usepackage[usenames]{color}
\usepackage{array}

\usepackage[
      colorlinks=true,
      linkcolor=blue,
      urlcolor=blue,
      filecolor=blue,
      citecolor=red,
      pdfstartview=FitV,
      pdftitle={},
      pdfauthor={},
      pdfsubject={},
      pdfkeywords={},
      pdfpagemode=None,
      bookmarksopen=true
]{hyperref}

\usepackage{epsfig}

\usepackage{hyperref}

\def\beq{\begin{eqnarray}}
\def\eeq{\end{eqnarray}}

\def \RR{{\mathbb{R}}}

\def\be{\begin{equation}}
\def\ee{\end{equation}}
\def\bea{\begin{eqnarray}}
\def\eea{\end{eqnarray}}

\def\be{\begin{equation}}
\def\ee{\end{equation}}
\def\bea{\begin{eqnarray}}
\def\eea{\end{eqnarray}}

\newcommand{\rom}[1]{\mathrm{#1}}

\def\cA{\mathcal{A}}

\def\cF{\mathcal{F}}

\def\cL{\mathcal{L}}
\def\cM{\mathcal{M}}

\def\cQ{\mathcal{Q}}

\def\cV{\mathcal{V}}

\def\nn{\nonumber}

\numberwithin{equation}{section}

\begin{document}

\begin{flushright}
\texttt{\today}
\end{flushright}

\begin{centering}

\vspace{2cm}

\textbf{\Large{Geroch Group Description of Black Holes}}

 \vspace{0.8cm}

  {\large
  Bidisha Chakrabarty and Amitabh Virmani}

  \vspace{0.5cm}

\begin{minipage}{.9\textwidth}\small
\begin{center}
{\it Institute of Physics,
Sachivalaya Marg, Bhubaneswar, Odisha, India 751005} \\
  \vspace{0.5cm}
{\tt bidisha, virmani@iopb.res.in}
\\ $ \, $ \\

\end{center}
\end{minipage}


\begin{abstract}
On one hand the Geroch group allows one to associate spacetime independent matrices with gravitational configurations that effectively only depend on two coordinates. This class includes stationary axisymmetric four- and five-dimensional black holes. On the other hand, a recently developed inverse scattering method allows one to factorize these matrices
to  explicitly construct the corresponding spacetime configurations. In this work we demonstrate the construction as well as the factorization of Geroch group matrices for a wide class of black hole examples. In particular, we obtain the Geroch group SL$(3,\RR)$ matrices for the five-dimensional Myers-Perry and Kaluza-Klein black holes and the Geroch group SU$(2,1)$ matrix for the four-dimensional Kerr-Newman black hole. We also present certain non-trivial relations between the Geroch group matrices and charge matrices for these black holes.
\end{abstract}

\end{centering}

\newpage

\tableofcontents


\setcounter{equation}{0}
\section{Introduction}
Solutions of a gravity theory coupled to matter in $d$ dimensions admitting $k$ commuting Killing vectors can be thought of as solutions of dimensionally reduced gravity theory in $d-k$ dimensions. Quite often the dimensionally reduced gravity theory admits an enhanced group of symmetries \cite{Julia, Breitenlohner:1987dg}, sometimes called the hidden symmetries. These hidden symmetries have been fruitfully used to study solutions of higher-dimensional theories for several decades now. Most notably, the hidden symmetry groups in three-dimensions have been used to construct black hole solutions of four- and five-dimensional theories, see, e.g., the review \cite{Youm:1997hw}, and references \cite{Chow:2013tia, Chow:2014cca} for recent works. They have also been used to obtain uniqueness results for four- and five-dimensional black holes, see, e.g., \cite{Tomizawa:2009ua, Hollands:2012cc}. More recently, these symmetry groups have been used to classify BPS and non-BPS solutions of various four-dimensional supergravity theories, see, e.g., the following incomplete list of references \cite{Bossard:2009at, Bossard:2009we, Bossard:2011kz, Bossard:2012ge, Bossard:2014yta}.

The case of $d-2$ commuting Killing vectors is particularly rich, as in that case the hidden symmetry groups are typically infinite dimensional Lie groups \cite{Geroch1970,Julia1981, BM, BMnotes, Nicolai}. We call the two dimensional hidden symmetry groups Geroch groups, by extension of the case of pure gravity in four-dimensions\footnote{In the 1990s many authors contributed to the development of Geroch group as symmetries of string theory, see e.g., the following incomplete list of references \cite{Bakas1, Sen, Bakas2, JM}.}.
The corresponding Lie algebras are the affine-extensions of the Lie algebras of the hidden symmetry groups in three dimensions. This is because, the two-dimensional models are often obtained via dimensional reduction from three dimensions over yet another Killing vector. This results in integrable models \cite{BM, BMnotes, BZ1, BZ2, BV}.

In this paper we consider black holes of five-dimensional vacuum gravity and four-dimensional Einstein-Maxwell theory and study them from the Geroch group perspective. We restrict our attention to three examples:  $(i)$ dyonic Kaluza-Klein black hole \cite{Rasheed:1995zv, Larsen:1999pp}, $(ii)$ dyonic Kerr-Newman black hole, and $(iii)$ five-dimensional doubly spinning Myers-Perry black hole \cite{Myers:1986un}. Using these three examples we exhibit the construction of Geroch group matrices. We obtain  Geroch group SL$(3,\RR)$ matrices for the Myers-Perry and Kaluza-Klein black holes and an SU$(2,1)$ matrix for the Kerr-Newman black hole. Along the way, we also present certain non-trivial relations between the Geroch group matrices and the corresponding charge matrices.

The motivation for studying these issues is manifold. Apart from identifying the precise Geroch group matrices for certain black holes, the examples worked out in this paper teach us more about the inverse scattering method recently proposed in \cite{Katsimpouri:2012ky, Katsimpouri:2013wka}. The  method proposed there is based on the Geroch group and it requires one to factorize Geroch group matrices in a certain way. In this paper we factorize the matrices for the examples mentioned in the previous paragraph. Our present study brings in two new elements: $(i)$ we extend the factorization algorithm developed there to incorporate five-dimensional asymptotically flat boundary conditions, $(ii)$ we present a fairly non-trivial example involving the group SU$(2, 1)$ of the general factorization algorithm presented there.

The rest of the paper is organized as follows. In section \ref{review2d} we start with a brief review of dimensional reduction to two dimensions, focusing on details that are most relevant for the rest of the paper. In section \ref{recipe} we present a simple and quite general recipe for computing the Geroch group matrix for a spacetime specified through a three-dimensional coset representative. In section \ref{GeneralConsiderations} we present certain general results on the Geroch group matrices and in section \ref{examples} we present explicit examples for the black holes mentioned above. We close with a summary and a brief discussion of open problems in section \ref{disc}. Certain technical details regarding SL$(3, \RR)$ and SU$(2,1)$ coset models are given in appendix \ref{appendix}. The appendix is an important part of the paper.

\section{Preliminaries: dimensional reduction to two dimensions}
\label{review2d}

In this section we present a brief review of dimensional reduction to two dimensions. We closely follow the notation and discussion of \cite{BM, Nicolai, Katsimpouri:2012ky, Katsimpouri:2013wka}.

We perform dimensional reduction of a higher-dimensional gravity theory to two dimensions in two steps. In the first step we reduce the theory to three-dimensions and in the second step we reduce it from three to two dimensions. We work with gravity matter systems that have some global symmetry G and some local symmetry K in three dimensions.  K is a maximal subgroup of G. A general element $k$  of the subgroup K satisfies $k^\sharp k = 1$, where hash $(\sharp)$ is the anti-involution that defines the coset G/K. Let $V(x)$ be a coset representative of G/K. We use $x$ to collectively denote the three coordinates of the three-dimensional space. The symmetries act on $V(x)$ as
\be
V(x) \to k(x) V(x) g, \label{leftaction}
\ee
with a global $g \in $ G and a local $k \in$ K. A more convenient object to work with is, see, e.g.~\cite{Pope},
\be
M(x) = V^\sharp(x) V(x),
\ee
with symmetries acting on $M(x)$  as\footnote{Following standard references,
see e.g., \cite{Pope}, we use the notation G$/$K instead of K$\backslash$G even
though we define the coset element $V(x)$ using a left action of K in equation
\eqref{leftaction}.
}
\be
M(x) \to g^\sharp M(x) g.
\ee

We next consider dimensional reduction over a spacelike Killing vector to two dimensions. In order to do so we write the three-dimensional metric as
\be
ds^2_3 = f^2(d \rho^2 + dz^2) + \rho^2 d\varphi^2.
\ee
These coordinates are called the Weyl canonical coordinates. The Killing vector $\partial_\varphi$ allows us to reduce the theory from three to two dimensions. The function $f$ multiplying the flat two-dimensional base metric $(d \rho^2 + dz^2)$ is called the conformal factor.  The resulting two-dimensional gravitational system
upon dimensional reduction along $\varphi$ direction is integrable, meaning that there exist a Lax pair whose compatibility condition is exactly the equations of the two-dimensional gravitational system. The quantity that one solves for in the Lax equations depends on a spectral parameter. There are several Lax formulations that one can write for the  two-dimensional gravitational system of interest. In this paper we will exclusively work with the Breitenlohner-Maison (BM) Lax pair \cite{BM, Nicolai}. It takes the following form in the notation of \cite{Katsimpouri:2012ky,Katsimpouri:2013wka}
\be
\partial_m \cV \cV^{-1} = Q_m + \frac{1-t^2}{{1+t^2}} P_m - \frac{2t}{1+t^2} \epsilon_{mn} P^n.
\label{Lax}
\ee
Here (i) we use the notation $x^m = (\rho,z)$ and from now onwards use $x$ to collectively denote the two-dimensional coordinates, (ii) the Lax equations require us to consider the generalization $V(x) \rightarrow \cV(t, x)$, a quantity that depends on the spectral parameter $t$ with the property $\cV(0,x) = V(x)$, (iii) $P_m$ and $Q_m$ are respectively the symmetric and anti-symmetric parts of the Lie algebra element $\partial_m V V^{-1} = P_m + Q_m$, $P_m^\sharp = P_m$ and $Q_m^\sharp = -Q_m$.

The integrability condition for equations \eqref{Lax} is equivalent to the equations of the motion of the two-dimensional gravitational system if and only if the spectral parameter satisfies certain spacetime dependent differential equation. That differential equation can be integrated to give
\be
t_\pm(w,x) =\frac{1}{\rho}\left[(z-w) \pm \sqrt{(z-w)^2 + \rho^2}\right] = - \frac{1}{t_\mp}(w,x).
\label{tpm}
\ee
The parameter $w$ is an integration constant and we will refer it to as the spacetime independent spectral parameter. Equation \eqref{tpm} defines a two-sheeted Riemann surface over the two-dimensional base space. We take the positive sign in \eqref{tpm} as the physical sheet. Whenever we write $t$ we mean $t_+$.

The anti-involution extends to functions $\cV(t,x)$ as
\be
(\cV(t,x))^\sharp = \cV^\sharp\left(-\frac{1}{t},x\right).
\ee
Similar to
$
M(x) = V^\sharp(x) V(x),
$
we can construct the so-called monodromy matrix,
\be
\cM(t,x) = \cV^\sharp\left(-\frac{1}{t},x\right) \cV(t,x).
\ee
A priori it appears that the matrix $\cM(t,x)$ is spacetime dependent, however, remarkably, using  the Lax equations, one can show that $\cM(t,x)$ is spacetime \emph{independent} \cite{BM, Nicolai}:
\be
\cM(t,x) = \cM(w).
\ee
It only depends on the spacetime independent spectral parameter $w$.  We call the monodromy matrix $\cM(w)$ the Geroch group matrix corresponding to the spacetime configuration described by $V(x)$. Thus, the Geroch group allows one to associate a spacetime independent matrix to a spacetime configuration that effectively depends on only two coordinates.

\section{Relation between $M(x)$ and $\cM(w)$}
\label{recipe}
The discussion of the previous section in principle allows one to associate a Geroch group matrix $\cM(w)$ to an arbitrary spacetime configuration described by $V(x)$. However, in order to do that one must first solve for $\cV(t,x)$ via the Lax equation \eqref{Lax}. This step may not be always easy.  Fortunately, one can arrive at a simple and quite general relation between the matrices $M(x)$ and $\cM(w)$. This relation allows one to construct  $\cM(w)$ rather directly from $M(x)$. In the following we first present this relation and then present a derivation of it.

Consider the two-dimensional space spanned by the canonical coordinates $(\rho, z)$. In the literature this space is sometimes known as the factor space. In reference \cite{Hollands:2007aj} Hollands and Yazadjiev studied the global structure of the factor space.  They showed that for spacetimes containing non-extremal
horizons, the corresponding factor space  is a manifold possessing a connected boundary with corners. The boundary is at $\rho=0$ and consists of a union
of intervals. The intervals either correspond to the horizon(s) or to the fixed points of the rotational Killing
vectors. The corners are the points where two adjacent intervals meet. We consider the cases where we can draw a semicircle of sufficiently large radius $R$ in the $(\rho, z)$ half-plane, such that all corners are inside this semicircle. With these assumptions the relation between $M(x)$ and $\cM(w)$ is simply
\be
M(\rho =0, z = w \mbox{\: with \:} z < - R) = \cM(w). \label{MzMw}
\ee
In this rest of this section we present a derivation of equation \eqref{MzMw} following \cite{BM}.

Let us start by noting that the spacetime independent spectral parameter $w$ in general can take complex values. However, in order to relate to the discussion of the previous paragraph, in particular for the replacement $z = w$ in  equation \eqref{MzMw} to make sense, we must take it to be real.  It turns out that the discussion below is better presented by taking $w$ to be complex to start with and  taking the limit $\mathbf{Im}(w) \to 0$ towards the end.

The functions $t_\pm$ defined via equation \eqref{tpm} considered as functions of complex $w$ have two branch points at
$
\rho = \pm \mathbf{Im}(w),  z = \mathbf{Re}(w).
$
At these branch points
\begin{align}
t_\pm \big{|}_{\rho =  \mathbf{Im}(w), \:  z = \mathbf{Re}(w)} & = - i &
t_\pm \big{|}_{\rho =  -\mathbf{Im}(w), \:  z = \mathbf{Re}(w)} &= + i.
\label{notuseful}
\end{align}
Note in particular that at the branch points $t_\pm$ take the same values and are consistently related by $t_\pm \to - \frac{1}{t_\mp}$ relation.

As mentioned above, for most of the consideration we take $t$ to mean $t_+$, i.e., when we write $\cV(t,\rho,z)$ we mean $\cV(t_+,\rho,z)$. However, for the discussion below we need to be more careful about the two-sheets of the Riemann surface defined by \eqref{tpm}, so we introduce one more notation
\be
 \cV_\pm(w,\rho,z) =  \cV(t_\pm(w, \rho, z),\rho,z).
\ee

Let us now concentrate on the region $\rho \to 0 $ and $z < - R$. In this region $t_+ \to 0$ and $t_- \to \infty$, as a result the Lax equations \eqref{Lax} in this region simplify to
\begin{align}
\partial  \cV_+  \cV_+^{-1} &= \partial  V  V^{-1}, & \partial  \cV_-  \cV_-^{-1} &= -(\partial  V  V^{-1})^\sharp.
\end{align}
These equations have simple solutions with the required property for $\cV_+$,
\bea
 \cV_+ (w, 0, z) &=&  V(0, z), \label{Vplus}\\
 \cV_-(w, 0, z) &=&  (V^\sharp(0,z))^{-1} C(w), \label{Vminus}
\eea
for some constant matrix $C(w)$. Since $t_\pm$ have the same values at the branch points, it follows that the functions $ \cV(t_\pm(w,\rho,z),  \rho, z)$ also have the same values at the branch points, i.e.,
\be
 \cV_+(w, \rho, z)\big{|}_{\rho =  \mathbf{Im}(w), \: z = \mathbf{Re}(w),} =  \cV_-(w, \rho, z)\big{|}_{\rho =  \mathbf{Im}(w), \: z = \mathbf{Re}(w)}.
\ee
In the limit $\mathbf{Im}(w) \to 0$, this implies
\be
 \cV_+(w, 0, w) =  \cV_-(w, 0,w). \label{VplusVminus}
\ee
Using relations \eqref{Vplus} and \eqref{Vminus} in \eqref{VplusVminus} we thus have
\be
 V(0,w) = (V^\sharp(0,w))^{-1} C(w).
\ee
Therefore,
\bea
 C(w) &=&  V^\sharp(0,w)  V(0,w) \\
 &=&  M(0,w).
\eea
Hence it follows that
\bea
 \cM(w) &=&  \cV^\sharp_-(w,0,z) \cV_+(w,0,z), \\
 &=& \left( (V^\sharp(0,z))^{-1} M(0,w) \right)^\sharp V(0, z),\\
 &=&  M(0,w).
\eea
This is precisely equation \eqref{MzMw}. To summarise, one can calculate the Geroch group matrix corresponding to an axisymmetric stationary space-time configuration by simply evaluating the matrix $M(x)$ in canonical coordinates at $\rho = 0$ and $z = w  \mbox{\: with \:} z < - R$. This is the recipe we use in the later sections to study Geroch group description of black holes.

\section{Geroch group matrices: general considerations}
\label{GeneralConsiderations}
In reference \cite{Katsimpouri:2012ky} Riemann-Hilbert factorization for SL(2) Geroch Group matrices was studied. This was later generalized to other groups, in particular to the case of SO(4,4) relevant for the so-called STU supergravity, in reference \cite{Katsimpouri:2013wka}. In both these studies attention was focused on four-dimensional asymptotically flat boundary conditions. In this paper, among other things, we generalize those studies to incorporate five-dimensional asymptotically flat boundary conditions. These boundary conditions bring in some minor changes to the factorization algorithm developed in \cite{Katsimpouri:2012ky, Katsimpouri:2013wka}. For simplicity, in this section we restrict our attention to SL(3,$\RR$) --- hidden symmetry group of vacuum five-dimensional gravity.

\subsection{Boundary conditions}
\label{BoundaryConditions}
In an interesting paper \cite{Giusto:2007fx} Giusto and Saxena pointed out that if dimensional reduction of  five-dimensional Minkowski space is done over appropriately chosen Killing vectors then the asymptotic limit of the coset matrix $M(x)$ is a constant matrix $Y$. The $Y$ matrix is different from the identity matrix. They also identified an SO(2,1) subgroup of SL(3) that leaves the constant matrix $Y$ invariant. It follows that all five-dimensional asymptotically flat solutions can be dimensionally reduced to three dimensions in a manner that asymptotically $M(x)$ is the constant matrix $Y$.  Using these inputs, in this paper we explore SL(3) Geroch group matrices and their factorization, where they asymptote to the constant matrix $Y$ different from the identity.
As in \cite{Katsimpouri:2012ky, Katsimpouri:2013wka} we restrict our attention to the so-called soliton sector. A general such matrix is of the form
\be
\cM(w) = Y + \sum_{k=1}^{N} \frac{A_k}{w-w_k}, \qquad \qquad A_k = \alpha_k a_k a_k^\sharp.
\label{Mw}
\ee
In particular, we allow for only simple poles in $w$, and we take the rank of the residues at these poles to be one. The residue matrices are $(\sharp)-$symmetric: $A_k^\sharp = A_k$. We expect that this choice includes several solutions of physical interest. As we will see in the following, it certainly includes the rotating Myers-Perry black hole.

The $(\sharp)-$operation on vectors $a_k$ is defined as $a_k^\sharp = a_k^T \eta$, where  $\eta$  is the quadratic form preserved by the denominator SO(2,1) subgroup of the coset SL(3)/SO(2,1).  In the following we focus our attention to the case when the dimensional reduction from five to three dimensions is done first over a spacelike direction and then over a timelike direction. In that case it follows from equation \eqref{gentranspose} of appendix \ref{appendix} that
\be
\eta = \mathrm{diag}\{1, -1, 1\}.
\ee
Furthermore, we restrict ourselves to the cases where the inverse of $\cM(w)$ also has poles at the same locations as $\cM(w)$ with residues of rank-one. We parameterize this matrix as
\be
\cM(w)^{-1} = Y^{-1} - \sum_{k=1}^{N} \frac{B_k}{w-w_k}, \qquad \qquad B_k = \beta_k \eta b_k b_k^T.
\ee
For later convenience we have put minus signs in front of the residues and we have put the $\eta$ matrix in the residues on the left hand side. $B_k$ matrices are also $(\sharp)-$symmetric: $B_k^\sharp = B_k$.

To find the explicit form of the matrix $Y$ we follow the steps of Giusto and Saxena. For five-dimensional Minkowski space in coordinates
\be
ds^2 = -dt^2 + dr^2 + r^2 (d\theta^2 + \sin^2\theta d\phi^2 +\cos^2\theta d\psi^2),
\ee
we define new coordinates
\bea
\phi_+ &=& \ell(\psi + \phi), \\
\phi_- &=& (\psi - \phi),
\eea
where $\ell$ is some arbitrary length scale.  Upon Kaluza-Klein reduction first along $ \phi_+$  and then along $t$ the resulting matrix $M(x)$
takes the form
\be
M(x) =\left(
 \begin{array}{ccc}
 \frac{4 \ell^2}{r^2} & 1 & 0 \\
 -1 & 0 & 0 \\
 0 & 0 & 1
\end{array}
\right),
\ee
and the three-dimensional base metric takes the form
\be
ds^2_3 = \frac{r^2}{4 \ell^2} \left( dr^2 + r^2 d\theta^2 + r^2 \sin^2 \theta \cos^2 \theta d\phi_-^2 \right).
\ee
Clearly, the matrix $M(x)$ asymptote to a constant matrix
\be
M(x) = Y + \mathcal{O}\left(\frac{1}{r^2}\right), \qquad  \qquad \mbox{with} \qquad \qquad
Y =
\left(
\begin{array}{ccc}
 0 & 1 & 0 \\
 -1 & 0 & 0 \\
 0 & 0 & 1
\end{array}
\right).
\label{Ymatrix}
\ee
The $Y$ matrix is symmetric under generalized transposition $Y^\sharp = Y$.

In the rest of this section we briefly present the changes $Y$ matrix brings in to the factorization algorithm of \cite{Katsimpouri:2012ky, Katsimpouri:2013wka}. Having specified the ansatz for $\cM(w)$ and $\cM^{-1}(w)$, we express $\frac{1}{w-w_k}$ in terms of the spacetime dependent spectral parameter $t$:
\begin{align}
\frac{1}{w-w_k} &= \nu_k \left( \frac{t_k}{t-t_k}+ \frac{1}{1+t t_k}\right), &
 & \mbox{where} &
 \nu_k &= -\frac{2 t_k}{\rho\left(1 + t_k^2 \right)},
\end{align}
and where the poles $t_k$ are determined by equation \eqref{tpm} evaluated at $w=w_k$ with the plus sign, $t_k = t_+(w=w_k)$. We wish to factorize $\cM(w)$ as
\begin{align}
\cM(w) = A_{-}^\sharp(t,x) M(x) A_{+}(t,x),
\label{factorize}
\end{align}
with $A_{-}(t,x) = A_{+} \left(-\frac{1}{t}, x \right)$ and $M^\sharp (x)= M(x)$. We make an ansatz for $A_+(t)$ and $A^{-1}_+(t)$
\begin{align}
A_+(t) &= \mathbf{I} - \sum_{k=1}^N \frac{t c_k a_k^T \eta}{1+t t_k}, \\
A^{-1}_+(t) &= \mathbf{I} + \sum_{k=1}^N \frac{t \eta b_k  d^T_k}{1+tt_k}.
\end{align}
and study the pole structure of the various matrix products to determine the vectors $c_k$ and $d_k$. This part of the analysis proceeds exactly as in \cite{Katsimpouri:2012ky, Katsimpouri:2013wka}, so we do not repeat it here. We present the results as a recipe.

The vectors $a_k$ and $b_k$ must satisfy
$
a^T_k \eta b_k = 0
$
for all $k$. The vectors $c_k$ and $d_k$ are determined from $a_k$ and $b_k$ by the matrix equations
\bea
c \Gamma &=& \eta b \\
\Gamma d &=& \eta a
\eea
where the matrix $\Gamma_{kl}$ is
\begin{align}
\label{GammaDef}
\Gamma_{kl} = \left\{ \begin{array}{ll}
 \frac{\gamma_k}{t_k} &\mbox{\qquad for \qquad $k=l$} \\
  \frac{1}{t_k-t_l} a_k^T b_l &\mbox{\qquad for \qquad $k \neq l$.}
       \end{array} \right.
\end{align}
The parameters $\gamma_k$ appearing in the $\Gamma$-matrix are determined by solving the equations
\begin{align}
a_k^T \eta \cA^{k} = \gamma_k \nu_k \beta_k b_k^T \qquad \mbox{and} \qquad
\cA_{k} \eta b_k = \gamma_k \alpha_k \nu_k a_k, \label{nosinglepole2}
\end{align}
with the definitions
\begin{align}
\cA^{k} =\left[ \cM^{-1}(t,x) + \frac{\nu_k \eta  \beta_k b_k
b_{k}^T }{1 + t t_k} \right]_{t=-\frac{1}{t_k}},
\qquad
\cA_{k} = \left[\cM(t,x) - \frac{\nu_k \alpha_k a_k   a_k^T \eta}{1 + t t_k} \right]_{t=-\frac{1}{t_k}}.
\end{align}
Now taking the limit $w \to \infty$ in \eqref{factorize} we find $Y = M(x) A_{+}(\infty,x)$, i.e,
\be
M(x) = Y  A^{-1}_{+}(\infty,x) = Y + Y t_k^{-1}\eta b_k (\Gamma^{-1})_{kl}a_l^T \eta.
\label{factorizationY}
\ee

\subsection{Two-soliton matrices}

So far we have only presented the general form of $\cM(w)$ and $\cM(w)^{-1}$ matrices. For arbitrary choices of the residue vectors $a_k, b_k$ and the parameters $\alpha_k, \beta_k$, these matrices do not belong to the group SL(3). The idea of using parameters $\alpha_k$ and $\beta_k$ is that by tuning them appropriately, various coset constraints can be imposed. In the case when there are only two poles in $\cM(w)$, it is relatively straightforward to take the coset constraints into account \cite{BMnotes}.

Let us choose the location of these poles to be $w_1 = +c$ and $w_2 = -c$. Let us take the components of the residue vectors $a_1$ and $a_2$ to be arbitrary. We introduce the notation $a = (a_1 \ \ a_2)$ where $a_1$ and $a_2$ are put as column vectors in a $3 \times 2$ matrix $a$.  Next consider the $2 \times 2$ matrix
\be
\xi  = a^T \eta Y^{-1} a =
\left(
\begin{array}{cc}
 a_1^T \eta Y^{-1}a_1 &  a_1^T \eta Y^{-1}a_2 \\
  a_2^T \eta Y^{-1}a_1 &  a_2^T \eta Y^{-1}a_2
\end{array}
\right).
\label{xi}
\ee
We note that since $Y$ is symmetric under generalized transposition, the matrix $\eta Y^{-1}$ is symmetric under the usual matrix transposition
\be
(\eta Y^{-1})^T = \eta Y^{-1}.
\ee
As a result the matrix $\xi$ defined in equation \eqref{xi} is a symmetric matrix. This crucial property allows us to choose $\alpha_1, \alpha_2, \beta_1, \beta_2$ and $b_1, b_2$ vectors in such a way that all coset constraints are satisfied:
\begin{align}
\alpha_1 &= \frac{2c}{\det \xi} \xi_{22}, & \alpha_2 &= -\frac{2c}{\det \xi} \xi_{11}, \\
\beta_1 &= - \frac{1}{\det \xi} \alpha_1,  & \beta_2 &= - \frac{1}{\det \xi} \alpha_2, &
\end{align}
and
\begin{align}
b &= (\det \xi) \eta Y^{-1} a  \xi^{-1} \epsilon,&
\epsilon &=
\left(
\begin{array}{cc}
0 & -1 \\
1 & 0
\end{array}
\right).
\end{align}
By construction $b$ is a $3 \times 2$ matrix whose columns are $b_1$ and $b_2$ vectors respectively. Of course there is an ambiguity in the normalization of these vectors. We have made a convenient choice in writing the above equations. Note that
\be
a^T b = (\det \xi) \epsilon.
\ee

\section{Geroch group matrices for black holes: examples}
\label{examples}
In this section we give explicit expressions for Geroch group matrices for a class of black hole examples. This section should be read in conjunction with appendix \ref{appendix}.
\subsection{Dyonic Kaluza-Klein}
Our first example is the dyonic Kaluza-Klein black hole  \cite{Rasheed:1995zv,
Larsen:1999pp}. The Kaluza-Klein black hole is conveniently written in terms of
four parameters $q, p, m, a$. These parameters respectively correspond to
electric and magnetic Kaluza-Klein charges, mass, and angular momentum.  For the discussion below we use exactly the form of the solution as given in reference \cite{Emparan:2007en} (in appendix A.1), except we use $a$ instead of $\alpha$ ($a_\rom{here} = \alpha_\rom{there}$) and $x$ instead of the polar angle $\theta$ on the two-sphere. The two are simply related by $x = \cos \theta$.
Using those expressions we construct three-dimensional scalars and from there the matrix $M(x)$. Expressions for the scalars and matrix $M(x)$ are somewhat lengthy, but after certain amount of manipulations, can be written in the following simpler form
\be
M(x) = g^\sharp M_\rom{Kerr}(x) g, \label{dyonic}
\ee
where
\be
g=\frac{1}{2 \sqrt{2} m}\left(
\begin{array}{ccc}
\sqrt{\frac{p (p+2m) (q+2m)}{p+q}} & \sqrt{(p-2 m) (q+2m)} & -\sqrt{\frac{q(p-2 m)(q-2 m)}{p+q}} \\
2\sqrt{\frac{q(p^2-4 m^2)}{p+q}} &  \sqrt{2p q} & -2  \sqrt{\frac{p(q^2-4 m^2)}{p+q}} \\
-\sqrt{\frac{p(p-2 m)(q-2 m)}{p+q}}
 & -\sqrt{(q-2 m)(p+2m)} &
\sqrt{\frac{p (p+2m) (q+2m)}{p+q}}
\end{array}
\right),
\label{GroupElementNiceForm}
\ee
with $g^\sharp = g^{-1}$, and where $M_\rom{Kerr}(x)$ is the matrix $M(x)$ for the Kerr solution,
\be
M_\rom{Kerr}(x) =
\left(
\begin{array}{ccc}
1+ \frac{2 m r}{r^2-2 m r+a^2 x^2} & 0 & -\frac{2 a m x}{r^2-2 m r+a^2 x^2} \\
 0 & 1 & 0 \\
 -\frac{2 a m x}{r^2-2 m r+a^2 x^2} & 0 & 1 + \frac{2 m (2 m-r)}{r^2-2 m r+a^2 x^2}
\end{array}
\right).
\ee

The first thing to note from the above expressions is the fact that the matrix $M(x)$ in equation \eqref{dyonic} for the rotating dyonic black hole is written as an action of an appropriate group element on the corresponding matrix for the Kerr solution $M_\rom{Kerr}(x)$. In fact, this is how the dyonic Kaluza-Klein black holes were constructed in the first place \cite{Rasheed:1995zv, Larsen:1999pp}. Expression for the group element $g$ in terms of Lie algebra generators  \eqref{kgenerators}
belonging to the denominator subgroup SO(2,1)
is as follows
\be
g = \exp(-\gamma k_3) \cdot \exp (-\beta k_1) \cdot \exp (\alpha k_2).
\ee
In this group element the generator $(\alpha k_2)$ generates KK magnetic charge and the generator $(-\beta k_1)$
generates KK electric charge. The generator $(-\gamma k_3)$ generates the four-dimensional Lorentzian NUT charge. The necessity
of acting with the generator $(-\gamma k_3)$ lies in the fact that the group element
$\exp (-\beta k_1) \cdot \exp (\alpha k_2)$
in addition to generating KK electric and magnetic charges also generates a four-dimensional  NUT charge. The final NUT charge
can be cancelled by appropriately tuning these parameters. To achieve this cancellation, we first use the
relation
\be
\tan 2 \gamma  = \tanh \alpha \sinh \beta,
\ee
and then the following somewhat unwieldy relations to write the group element in the form \eqref{GroupElementNiceForm}
\begin{align}
\cos \gamma &= \frac{\sqrt{p + 2m}\sqrt{q + 2m}}{\sqrt{2}\sqrt{p q + 4 m^2}}, &
\coth \beta &= \sqrt{\frac{q}{p}}\sqrt{\frac{pq + 4m^2}{q^2 -4m^2}}.
\end{align}

In order to obtain the Geroch group matrix corresponding to the above matrix $M(x)$ we need to introduce the canonical coordinates. Examining the determinant of the metric components on the Killing directions we get the canonical coordinates,
\begin{align}
\rho^2 &= (r^2 + a^2 - 2m r) (1-x^2), & z&=(r-m) x.
\end{align}
Using these relations we can in principle write the matrix $M(x)$ in the canonical form, however, it is easier to first introduce the prolate spherical coordinates \cite{Harmark:2004rm}
\begin{align}
c^2 (u^2-1)(1-v^2)& =(r^2 + a^2 - 2m r) (1-x^2), & c u v &= (r- m) x, &c &= \sqrt{m^2 -a^2}.
\end{align}
These relations can be solved to give $u = \frac{1}{c}(r-m), v = x.$
The transformation from the prolate spherical coordinates $(u,v)$ to the canonical coordinates is
\begin{align}
&u = \frac{\sqrt{\rho^2 + (z+c)^2}+\sqrt{\rho^2 + (z-c)^2}}{2c}, & v = \frac{\sqrt{\rho^2 + (z+c)^2}-\sqrt{\rho^2 + (z-c)^2}}{2c}.
\end{align}
Given these expressions it is easy to see that the limit $\rho \to 0$ and taking $z$ near $-\infty$
 amounts to the replacement $u \to -\frac{z}{c}$ and $v \to -1$, equivalently
$r \to - z + m$ and $x \to -1$.
Making these replacements in the matrix $M(x)$ we find the monodromy matrix $\cM(w)$. It takes the form
\be
\cM(w) = \mathbf{I} + \frac{A_1}{w-c}+ \frac{A_2}{w+c},
\label{monodromyKK}
\ee
where $A_1 = \alpha_1 a_1 a_1^\sharp$, $A_2 = \alpha_2 a_2 a_2^\sharp$. The vectors $a_1$ and $a_2$ for the dyonic black hole are
\begin{align}
a_1 &= g^\sharp a_1^{\rom{Kerr}}, & a_1^{\rom{Kerr}} = \left\{\zeta, 0, 1 \right\}, \\
a_2 &= g^\sharp a_2^{\rom{Kerr}},  & a_2^{\rom{Kerr}} = \left\{1, 0, \zeta \right\},
\end{align}
and the  parameters $\alpha_1$ and $\alpha_2$ are
\begin{align}
\alpha_1 &= \frac{2c(1+\zeta^2)}{(1-\zeta^2)^2},& \alpha_2 & = - \frac{2c(1+\zeta^2)}{(1-\zeta^2)^2}.
\end{align}
where $\zeta = \frac{m - c}{a}.$ The $a \to 0$ limit is perfectly smooth: in this limit $c \to m$ and $\zeta \to 0$.

The Riemann-Hilbert factorization of the monodromy matrix \eqref{monodromyKK} give the dyonic Kaluza-Klein black hole, as expected. This factorization is similar to the examples considered in \cite{Katsimpouri:2012ky, Katsimpouri:2013wka}, so we skip the details.

We now write some relations between the vectors obtained above, the charge matrix for KK black hole, and the monodromy matrix. We concentrate on the Kerr black hole --- for the dyonic KK black hole the corresponding relations are simply obtained by conjugation with $g$. The charge matrix $\cQ$ for a four-dimensional asymptotically flat configuration is defined as \cite{Bossard:2009at}:
\be
M(x) = \mathbf{I} - \frac{\cQ}{r} + \mathcal{O}\left(\frac{1}{r^2} \right).
\ee
In our normalization we have
\be
\cQ = - 2m h_2, \label{cQKerr}
\ee
where $h_2$ is the Cartan generator defined in \eqref{basis}. The charge matrix \eqref{cQKerr} satisfies the characteristic equation
\be
\cQ^3 - \frac{1}{2} \mbox{Tr} (\cQ^2) \cQ = 0,
 \ee
with  $\mbox{Tr} (\cQ^2) = 8m^2.$ The asymptotic form of the matrix $\cM(w)$ is also determined by the charge matrix
\be
\cM(w) =\mathbf{I} + \frac{\cQ}{w} +  \mathcal{O}\left(\frac{1}{w^2}\right).
\ee
Hence, it follows that
\be
\cQ = \sum_{i=1}^{2} \alpha_i a_i^{\rom{Kerr}} (a_i^{\rom{Kerr}})^\sharp. \label{cQgeneral}
\ee
We want to emphasize that relation \eqref{cQgeneral} is in fact quite general. For the present case only two poles are present in the Geroch group matrix so the sum in \eqref{cQgeneral} runs over only two values of the indices. When the Geroch group matrix has $N$ poles, the charge matrix is simply the sum of the residues at the poles.

The charge matrix as defined above does not capture information about the angular momentum of the spacetime. To encode that we can introduce one more matrix (see also \cite{Andrianopoli:2013jra} for a related construction)\footnote{We thank Guillaume Bossard for discussions on these ideas.}
\be
\cA = 2 a (e_3 - e_3^\sharp),
\ee
where $a$ is the Kerr rotation parameter and $e_3$ is one of the raising generators defined in \eqref{basis}. This matrix allows us to write some useful relations. Firstly, we observe that it anticommutes with the charge matrix,
\be
\{\cQ, \cA \} =0.
\ee
Secondly, using this matrix we can write yet another characteristic equation that captures rotation properties also
\be
(\cQ + \cA)^3 - \frac{1}{2} \mbox{Tr}\left((\cQ+\cA)^2\right) (\cQ+\cA) = 0,
\label{chNew}
\ee
with  $\mbox{Tr} \left((\cQ+\cA)^2 \right) = 8(m^2-a^2).$ Finally, we can write the full Geroch matrix solely in terms of the $\cQ$ and $\cA$ matrices
\be
\cM(w) =  \mathbf{I} + \frac{1}{w^2-c^2} \left( w \cQ +\frac{1}{2} \cQ^2 - \frac{1}{4}[\cQ, \cA] \right).
\label{MwcQ}
\ee
Note that although the matrix $\cA$ belongs to the
invariant $\mathfrak{so}(2,1)$ subalgebra, only the commutator 
$[\cQ,\cA]$, which belongs to the complement of $\mathfrak{so}(2,1)$ in
$\mathfrak{sl}(3,\RR)$ enters the Geroch group matrix \eqref{MwcQ}.

\subsection{Dyonic Kerr-Newman}
\label{KerrNewman}
In this section we explore dyonic Kerr-Newman black hole as a solution of four-dimensional Einstein-Maxwell theory from the Geroch group perspective.
It is well known that the symmetry group of the dimensionally reduced Einstein-Maxwell theory is SU(2,1) \cite{Kinnersley}.
For the case of the timelike reduction to three-dimensions the relevant coset is
\be
\frac{\mathrm{SU}(2,1)}{\mathrm{SL}(2,\RR) \times \mathrm{U}(1)}.
\label{SUcoset}
\ee
A detailed construction of the coset model is presented in appendix \ref{appendixA2}.

The metric and vector for dyonic Kerr-Newman black hole are given in Boyer-Lindquist coordinates as
\bea
ds^2 &=& -\frac{\Delta - a^2 \sin^2 \theta}{\Sigma} dt^2 - 2 a \left( \frac{r^2 + a^2 - \Delta}{\Sigma} \right) \sin^2\theta dt d\phi + \frac{\Sigma}{\Delta} dr^2 + \Sigma d\theta^2 \nn
\\
& & + \left( \frac{(r^2 + a^2)^2 - \Delta a^2 \sin^2 \theta}{\Sigma}\right) \sin^2 \theta d\phi^2  \\
A &=& \frac{1}{\Sigma} \left[-q r + a p  \cos \theta \right] dt +   \frac{1}{\Sigma} \left[ -p \cos \theta (r^2 + a^2) +  a q r \sin^2\theta \right] d\phi
\eea
where
\bea
\Sigma &=& r^2 + a^2 \cos^2 \theta, \\
\Delta &=& r^2 + a^2 + q^2 + p^2 - 2 m r.
\eea
The configuration is parameterized by four parameters: mass parameter $m$,
rotation parameter $a$, and electric and magnetic charges $q$ and $p$
respectively. For computational convenience we use $x = \cos \theta$ instead of the azimuthal angle $\theta$ for the rest of the discussion.  The four coset scalars take the form (for precise definition of these scalars we refer the reader to appendix \ref{appendixA2})
\begin{align}
e^\phi & = \frac{\Sigma}{\Sigma + p^2 + q^2 -2 m r}, &
\psi &= \frac{\sqrt{2} a m x}{\Sigma}, &
\chi_e &= -\frac{q r - a p x}{\Sigma},&
\chi_m &= -\frac{ p r + a q x}{\Sigma},
\end{align}
and as a result the matrix $M(x)$ is
\bea
M(x) &=&
\frac{1}{r^2 + a^2 x^2 + p^2 + q^2 -2 m r} \times \label{MxKN}\\
&& \left(
\begin{array}{ccc}
 r^2+a^2 x^2 & -\sqrt{2} (i p+q) (r+i a x) & p^2+q^2+2 i a m x \\
 \sqrt{2} (q-i p) (r-i a x) & -p^2-q^2+r^2+a^2 x^2-2 m r & \sqrt{2} (q-i p) (2
   m-r+i a x) \\
 p^2+q^2-2 i a m x & \sqrt{2} (i p+q) (-2 m+r+i a x) & (r-2 m)^2+a^2 x^2
\end{array}
\right). \nonumber
\eea

It is not difficult to verify that this matrix belongs to the coset \eqref{SUcoset}. For this, we need to check that the matrix  belongs to the group SU$(2,1)$ and that it is symmetric under the appropriate generalized transposition. The SU$(2,1)$ property and the generalized transposition are respectively defined in \eqref{kappa} and \eqref{SUgentranspose} in the appendix. Indeed, both the conditions are satisfied.

Once again, we use relation \eqref{MzMw} to find the Geroch group matrix. For this we first need to determine the canonical coordinates. By examining the determinant of the metric components on the Killing directions we get the canonical coordinates,
\be
\rho^2 = \Delta (1-x^2), \qquad z=(r-m) x.
\ee
Using these relations one can in principle write the matrix $M(x)$ in the canonical coordinates, however, as before, it is useful to first write the matrix in the prolate spherical coordinates
\be
u = \frac{1}{c}(r-m), \qquad v = x, \qquad c = \sqrt{m^2 -a^2 -p^2 -q^2},
\ee
and then convert to the canonical coordinates.
After doing the appropriate replacements in $M(x)$ we find
\bea
&&\cM(w) =
\frac{1}{(w^2 - c^2)}  \times \label{MwKN} \\
&& \left(
\begin{array}{ccc}
a^2+(m-w)^2 & -\sqrt{2} (p-i q) (a+i (m-w)) & p^2+q^2-2 i a m \\
 \sqrt{2} (p+i q) (a-i (m-w)) & a^2-m^2-p^2-q^2+w^2 & -\sqrt{2} (p+i q) (a+i
   (m+w)) \\
 p^2+q^2+2 i a m & \sqrt{2} (p-i q) (a-i (m+w)) & a^2+(m+w)^2
\end{array}
\right). \nonumber
\eea
This spacetime independent matrix is sufficient to determine the full Kerr-Newman configuration.
Since the Riemann-Hilbert factorization for this example is a little different from the examples previously discussed in the literature, we present certain details.

To write expressions in a less cumbersome manner we need to introduce some additional notation. We parameterize charges as
\begin{align}
m &= \mu \cosh 2 \beta, &  q &= \mu \sinh 2 \beta \cos b, &  p &= \mu \sinh 2 \beta \sin b,
\end{align}
and choose vectors $a_1$ and $a_2$ as
\begin{align}
a_1 &= g a_1^\rom{Kerr}, & a_1^\rom{Kerr}  &= \left\{-\frac{i a}{c+\mu},0,1\right\},\\
a_2 &= g a_2^\rom{Kerr}, & a_2^\rom{Kerr} &= \left\{1,0,\frac{i a}{c+\mu}\right\},
\end{align}
where
\be
g=e^{\frac{i b}{3}} \left(
\begin{array}{ccc}
 c^2_\beta & \sqrt{2}  s_\beta c_\beta
    & s^2_\beta \\
 \sqrt{2} e^{-i b} s_\beta c_\beta & e^{-i b} (2 c^2_\beta - 1) & \sqrt{2} e^{-i b} s_\beta c_\beta \\
  s^2_\beta & \sqrt{2}  s_\beta c_\beta
  & c^2_\beta
\end{array}
\right).
\ee
In terms of these vectors, the matrix $\cM(w)$ in equation \eqref{MwKN} can be written in a more recognizable form
\be
\cM(w) = \mathbf{I} + \frac{A_1}{w-c} + \frac{A_2}{w+c},
\ee
with
\be
A_k = \alpha_k a_k a_k^\sharp =\alpha_k a_k a_k^\dagger \eta,
\ee
and
\begin{align}
\alpha_1 &= \mu \left(1 + \frac{\mu}{c}\right), & \alpha_2 &= - \mu \left(1 + \frac{\mu}{c}\right).
\end{align}
The inverse matrix is parameterized as\footnote{The matrix  $\kappa$ defines the SU$(2,1)$ property, see equation \eqref{kappa}.}
\be
\cM(w)^{-1} =  \kappa^{-1} \cM^\dagger(w) \kappa = \mathbf{I} - \frac{B_1}{w-c} - \frac{B_2}{w+c},
\ee
with
\be
B_k = (-\alpha_k) (\kappa \eta a_k) (\kappa \eta a_k)^\sharp.
\ee
We note that $\kappa^{-1} = \kappa$ and $\kappa \eta = \eta \kappa$. Now all expressions are in the notation of \cite{Katsimpouri:2013wka} and the factorization proceeds exactly as discussed there. Following those steps we recover the coset matrix of equation \eqref{MxKN}.

In this case one can also construct an $\cA$ matrix that anticommutes with the
charge matrix $\cQ$ and satisfies \eqref{chNew} and \eqref{MwcQ}.

\subsection{Five-dimensional Myers-Perry}
The metric components of the doubly rotating Myers-Perry (MP) black hole in our conventions are
\begin{align}
& g_{rr} = \frac{\Sigma r^2}{(r^2 + l_1^2)(r^2 + l_2^2)- 2m r^2},  & & g_{xx} = \frac{\Sigma}{1-x^2}, \\
& g_{tt} = - \frac{\Sigma - 2m}{\Sigma},  & & g_{\psi \phi} = \frac{2 m l_1 l_2 x^2 (1-x^2)}{\Sigma}, \\
& g_{t \phi} = - \frac{2 m l_1 (1-x^2)}{\Sigma}, & & g_{t\psi} = - \frac{2 m l_2 x^2}{\Sigma}, \\
& g_{\phi \phi} =   \frac{1-x^2}{\Sigma} \left((r^2 + l_1^2)\Sigma + 2 m l_1^2 (1-x^2)\right), &
& g_{\psi \psi} = \frac{x^2}{\Sigma} \left((r^2 + l_2^2)\Sigma + 2 m l_2^2 x^2\right),
\end{align}
where
\be
\Sigma = r^2 + l_1^2 x^2 + l_2^2 (1-x^2).
\ee
As in the previous examples, for computational convenience we use $x =\cos \theta$ instead of the polar angle $\theta$. Variables
 $m$, $ l_1$, and $l_2$ are the mass and the two rotation parameters respectively.

For reasons mentioned in section \ref{BoundaryConditions} we define
\bea
\phi_+ &=& \ell(\psi + \phi), \label{dimred} \\
\phi_- &=& (\psi - \phi),
\eea
and perform Kaluza-Klein reduction first along $ \phi_+$  and then along the $t$
direction. The resulting scalars are somewhat cumbersome. Appropriately choosing the axionic shifts for the scalars $\chi_1, \chi_2, \chi_3$ we indeed find that the resulting matrix $M(r,x)$ has the asymptotic behaviour
\be
M(r,x) = Y + \mathcal{O}\left( \frac{1}{r^2}\right),
\ee
where $Y$ is the constant matrix \eqref{Ymatrix}. In order to construct the monodromy matrix $\cM(w)$ from $M(r,x)$ we first need to change coordinates to the canonical coordinates $(\rho,z)$ and then take the limit $\rho \to 0$ and take $z$ near $-\infty$ and finally replace $z$ with $w$.

The relation between the coordinates used above and the canonical coordinates is \cite{Harmark:2004rm}
\be
\rho = r x \sqrt{\Delta (1-x^2)}, \qquad \qquad z = \frac{1}{2}r^2 \left( 1- \frac{2m - l_1^2- l_2^2}{2 r^2} \right) (2x^2 -1),
\ee
where $\Delta$ is
\be
\Delta = r^2 \left(1 + \frac{l_1^2}{r^2}\right)\left(1+ \frac{l_2^2}{r^2}\right)- 2m.
\ee
In practice, performing this change of coordinates is not easy. It is easier to first introduce the prolate spherical coordinates $(u,v)$ and then change to the canonical coordinates. The prolate spherical coordinates are defined via
\begin{align}
&\alpha^2 (u^2 -1)(1-v^2) = r^2 x^2 \Delta  (1-x^2), &\alpha u v = \frac{1}{2} r^2 \left( 1- \frac{2m - l_1^2- l_2^2}{2 r^2} \right) (2x^2 -1),
\end{align}
or equivalently
\begin{align}
u&= \frac{1}{4\alpha} (2 r^2 + l_1^2 + l_2^2 - 2 m ), & v &= 2 x^2 -1,  & \alpha &= \frac{1}{4}\sqrt{(2 m - l_1^2 - l_2^2)^2 - 4 l_1^2 l_2^2}.
\end{align}
The transformation from the prolate spherical coordinates to the canonical coordinates is
\begin{align}
&u = \frac{\sqrt{\rho^2 + (z+\alpha)^2}+\sqrt{\rho^2 + (z-\alpha)^2}}{2\alpha}, & v = \frac{\sqrt{\rho^2 + (z+\alpha)^2}-\sqrt{\rho^2 + (z-\alpha)^2}}{2\alpha}.
\end{align}
Given these expressions it is easy to see that the limit $\rho \to 0$ and taking $z$ near $-\infty$
 amounts to the replacement $u \to -\frac{z}{\alpha}$ and $v \to -1$, equivalently
$r^2 \to - 2 z - \frac{1}{2}(l_1^2 + l_2^2 - 2m)$ and $x \to 0$.

Making these replacements in the matrix $M(r,x)$  we find the monodromy matrix $\cM(w)$. It takes the form
\be
\cM(w) = Y + \frac{A_1}{w-\alpha}+ \frac{A_2}{w+\alpha},
\label{monodromyMP}
\ee
where $A_1 = \alpha_1 a_1 a_1^T \eta$, $A_2 = \alpha_2 a_2 a_2^T \eta$, and $\eta = \mbox{diag}\{1,-1,1\}$. In particular, the Myers-Perry monodromy matrix has two poles at locations $w = \pm \alpha$ and the residues at these poles are of rank one. Clearly there is an ambiguity in the choice of the $a$ vectors and the $\alpha$ parameters. We choose these quantities such that they have smooth limits when either of the rotation parameters $l_1$ or $l_2$ go to zero. An explicit form of the $a$ vectors and the parameters $\alpha$'s is as follows
\bea
a_1 &=&
\left\{-\frac{\ell \left(4 \alpha +l_1^2-l_2^2-2 m\right)}{2 l_2 m},\frac{4 \alpha +(l_1+l_2) (l_1+3 l_2)-2 m}{8 \ell l_2},1\right\}, \\
a_2 &=&
\left\{\frac{\ell \left(4 \alpha -l_1^2+l_2^2+2 m\right)}{2 \sqrt{2} m},\frac{-4 \alpha +(l_1+l_2) (l_1+3 l_2)-2 m}{8 \sqrt{2}
   \ell},\frac{l_2}{\sqrt{2}}\right\},\\
\alpha_1 &=&
\frac{m \left(l_1^4-4 \alpha  l_1^2-2 l_1^2 \left(l_2^2+2 m\right)+4 \alpha  l_2^2+\left(l_2^2-2 m\right)^2+8 \alpha  m\right)}{2
   \left((l_1-l_2)^2-2 m\right) \left((l_1+l_2)^2-2 m\right)},\\
\alpha_2 &=&
\frac{m \left(l_1^4+4 \alpha  l_1^2-2 l_1^2 \left(l_2^2+2 m\right)-4 \alpha  l_2^2+\left(l_2^2-2 m\right)^2-8 \alpha  m\right)}{l_2^2
   \left((l_1-l_2)^2-2 m\right) \left((l_1+l_2)^2-2 m\right)}.
\eea
 In the limit $l_1 \to 0$ these expressions simplify to
\begin{align}
\alpha_1 &= \frac{2 m^2}{2 m - l_2^2}, & a_1 &= \left\{\frac{\ell l_2}{m},\frac{l_2}{4 \ell},1\right\}, \\
\alpha_2 &= -\frac{2 m}{2 m - l_2^2}, & a_2 &= \left\{\sqrt{2} \ell ,\frac{l_2^2-m}{2 \sqrt{2} \ell},\frac{l_2}{\sqrt{2}}\right\}.
\end{align}
Additionally, in the limit $l_2 \to 0$ whereupon we obtain Schwarzschild black hole, these expressions further simplify to
\begin{align}
\alpha_1 &= m, & a_1 &= \left\{0,0,1\right\}, \\
\alpha_2 &= -1, & a_2 &= \left\{\sqrt{2} \ell,-\frac{m}{2 \sqrt{2} \ell},0\right\}.
\end{align}
These last expressions are especially informative. Since $\alpha_1$ is equal to $m$ and the vector $a_1$ is simply a constant, in the limit $m \to 0$  the residue of the pole $w = + \alpha$ vanishes, i.e., the pole $w = + \alpha$ disappears. Whereas the residue of the pole $w = - \alpha$ does not vanish in the same limit. In fact in this limit the pole location $\alpha$ also goes to zero. This limit corresponds to five-dimensional Minkowski space. The monodromy matrix simplifies to
\be
\cM(w) = Y + \frac{\alpha_2 a_2 a_2^T \eta}{w},
\ee
with $\alpha_2 = -1$ and  $a_2 = \left\{\sqrt{2} \ell,0,0\right\}$. Thus, from the Geroch group point of view five-dimensional Minkowski space has a non-trivial monodromy matrix. Its asymptotic limit is the constant matrix $Y$. It also has a pole at $w=0$ with residue of rank one. The residue depends on the parameter $\ell$ introduced via equation \eqref{dimred} and as such it can take any non-zero value.

The Riemann-Hilbert factorization of the monodromy matrix \eqref{monodromyMP} gives the double spinning Myers-Perry solution, as expected. Details of this factorization are also very similar to the examples considered in \cite{Katsimpouri:2012ky, Katsimpouri:2013wka}. The only difference is that one needs to take into account the $Y$ matrix via equation \eqref{factorizationY}.

We end this section by observing some properties of the monodromy matrix \eqref{monodromyMP} in relation to the charge matrix for the Myers-Perry spacetime.

The concept of charge matrix for the five-dimensional boundary conditions was introduced in \cite{Hornlund:2010tr}. In the present context it is computed as follows.
First we define
\be
D=\left(
\begin{array}{ccc}
 \frac{1}{\sqrt{2}} & \frac{1}{\sqrt{2}} & 0 \\
 \frac{1}{\sqrt{2}} & -\frac{1}{\sqrt{2}} & 0 \\
 0 & 0 & 1
\end{array}
\right),
\ee
with the property that
\be
D^\sharp D = Y.
\ee
Next we conjugate $M(r,x)$ with $D^{-1}$. This operation has the effect that the matrix
\be
(D^\sharp)^{-1} M(r,x)D^{-1}
\ee
asymptote to identity matrix. Then, the charge matrix is defined by the coefficient of $r^{-2}$ term in the asymptotic expansion near infinity
\be
(D^\sharp)^{-1} M(r,x)D^{-1} = \mathbf{I} - \frac{2\cQ}{r^2 } + \mathcal{O}\left(\frac{1}{r^4} \right).
\ee
The charge matrix obtained in this way satisfies the characteristic equation
\be
\cQ^3 - \frac{1}{2} \mbox{Tr} (\cQ^2) \cQ = 0.
\ee
It does not capture  information about both  angular momentum $l_1$ and $l_2$ of the MP spacetime. To encode that we can introduce one more matrix
\be
\cA = a_1 (e_1 - e_1^\sharp)+  a_2 (e_2 - e_2^\sharp) + a_3 (e_3 - e_3^\sharp),
\ee
with coefficients
\begin{align}
a_1 &= - \frac{m- 4 \ell^2}{4\sqrt{2}\ell}(l_1 - l_2),
&
a_2 &= -\frac{1}{2}(l_1 + l_2)(l_1 - l_2),
&
a_3 &= - \frac{m+ 4 \ell^2}{4\sqrt{2}\ell}(l_1 - l_2),
\end{align}
where $e_1, e_2, e_3$ are the raising generators defined in \eqref{basis}. This matrix allows us to write relations similar to the ones written above for dyonic KK black hole. Firstly, we observe that it anticommutes with the charge matrix,
\be
\{\cQ, \cA \} =0.
\ee
Secondly, using this matrix we can write yet another characteristic equation that captures rotation properties as well,
\be
(\cQ + \cA)^3 - \frac{1}{2} \mbox{Tr}\left((\cQ+\cA)^2\right) (\cQ+\cA) = 0,
 \ee
with  $\mbox{Tr} \left((\cQ+\cA)^2 \right) = 8\alpha^2$. Finally, we can write the full Geroch matrix solely in terms of $\cQ$, $\cA$, and $D$ matrices
\be
\cM(w) =  Y + \frac{1}{w^2-\alpha^2} D^\sharp \left( w \cQ +\frac{1}{2} \cQ^2 - \frac{1}{4}[\cQ, \cA] \right)D.
\ee

\section{Summary and open problems}
\label{disc}
In this paper we have analysed Geroch group description of black holes. We presented a general relation, equation \eqref{MzMw}, between the three-dimensional coset matrix $M(x)$ and the Geroch group matrix $\cM(w)$. Using this simple relation we constructed Geroch group matrices for dyonic Kaluza-Klein black hole, five-dimensional Myers-Perry black hole, and for Kerr-Newman black hole. Along the way, we presented some non-trivial relations between the Geroch group matrices and charge matrices. We also incorporated five-dimensional asymptotically flat boundary conditions in the factorization algorithm of \cite{Katsimpouri:2012ky, Katsimpouri:2013wka}.

There are several ways in which our study can be extended. Perhaps the simplest such  extension will be to work out similar details for the Einstein-Maxwell dilaton-axion model (EMDA) that has
the hidden symmetry group Sp$(4,\RR)$. Equally interesting is the case of bosonic sector of the $N=2$ supergravity with one vector multiplet with prepotential $F = - i X_0 X_1$. This theory has the hidden symmetry group SU$(2,2)$. In
both these cases we do not expect to meet any surprises. We expect that a straightforward extension of the above discussion will be applicable. Along the same lines the Geroch group SO(4,4) matrix for the five-dimensional Cveti{\v c}-Youm black hole \cite{CY5d} can also be obtained\footnote{The Geroch group SO(4,4) matrix for the four-dimensional Cveti{\v c}-Youm black hole \cite{Cvetic:1996kv} was obtained in \cite{Katsimpouri:2013wka}.}.

A theory that requires new ideas is minimal supergravity in five-dimensions. This theory has hidden symmetry group to be the smallest
exceptional group G$_{2(2)}$. For this set-up also, given $M(x)$ one can
construct $\cM(w)$ using \eqref{MzMw}.
The residues at the poles will turn out to be of rank-2. Since the residues
are of rank-2 one needs to separate out contributions into the $a-$ and
$b-$vectors. This seems to be a non-trivial step. Moreover, since the defining
relation for G$_{2(2)}$ matrices in fundamental representation is
non-linear, 
\be
c_{abc}\cM_{aa'}\cM_{bb'}\cM_{cc'} = c_{a'b'c'},
\ee
with $c_{abc}$ the $\mathfrak{g_{2(2)}}$ invariant three-form,
most likely certain details of the factorization algorithm as presented in \cite{Katsimpouri:2013wka} 
need to be adjusted (see also related comments in
\cite{Figueras:2009mc}). However, we do hope that working out examples of 
Geroch group matrices using \eqref{MzMw}, as we have done in this paper, will
shed some light on those issues as well.

For practical calculations involving more complicated solutions such as black rings and two-centered black holes we need to consider cases where Geroch group matrices need not approach a constant matrix at infinity. For example, for neutral S$^1$ rotating Emparan-Reall black ring, it is most manageable to do dimensional reduction first along the S$^1$ direction, and then along the time direction. The matrix $M(x)$ obtained that way does not asymptote to a constant matrix at spatial infinity, and for the same reason the matrix $\cM(w)$ also does not approach a constant matrix at infinity ($w=\infty$). In fact, the matrix $\cM(w)$ has a pole at $w=\infty$. The factorization algorithm developed in \cite{Katsimpouri:2012ky, Katsimpouri:2013wka} does not incorporate this feature.  It will be worthwhile to extend the previously developed factorization algorithms to allow for this possibility.  Such considerations will be
the natural arena for describing black rings and related set-ups from the Geroch group point of view. We hope to report on some of these issues in the future.

Finally, it will be very interesting to understand AdS boundary conditions from the Geroch group perspective. Some preliminary steps in this direction are taken in reference \cite{Leigh:2014dja}, though a lot remains to be understood.

\subsection*{Acknowledgements}
We are grateful to Despoina Katsimpouri  and Axel Kleinschmidt for useful discussions, and especially to Guillaume Bossard for useful discussions and for sharing some of his unpublished notes. AV would also like to thank AEI Potsdam, CPHT -- Theoretical Physics Center at Ecole Polytechnique at Saclay, and KITPC Beijing for their warm hospitality where part of this work was done.

\appendix

\section{Coset models}
\label{appendix}
In this appendix we present construction of relevant coset models. The discussion below is fairly standard, to set up our notation for the main text we present certain details.
\subsection{SL(3,\,$\RR$)/SO(2,1)}
\label{appendixA1}
Let us start with a discussion of SL(3,\,$\RR$)/SO(2,1) coset relevant for five-dimensional vacuum gravity.  The Lagrangian for vacuum gravity is
$ \mathcal{L}_5=R_5 \star 1.$
We perform KK reduction to three dimensions using the ansatz \cite{Pope}
\bea
ds^2_5 &=& e^{\frac{1}{\sqrt{3}}\phi_1+\phi_2}ds^2_3+\epsilon_2 e^{\frac{\phi_1}{\sqrt{3}}-\phi_2}\left(dz_4+\mathcal{A}_{(1)}^2\right)^2            +\epsilon_1e^{-\frac{2\phi_1}{\sqrt{3}}} \left(dz_5
+\chi_1dz_4+\mathcal{A}_{(1)}^1 \right)^2,
\label{ansatz}
\eea
where reduction is first done along $z_5$ and then along $z_4$. Here $\epsilon_1$ and $\epsilon_2$ take values $\pm 1$, they respectively denote the signature of the first and second direction over which reduction from five to three dimensions is performed. We will take one of them to be $-1$ and the other $+1$.

The reduced three-dimensional Lagrangian in terms of the fields  appearing in \eqref{ansatz} is
\bea
\mathcal{L}_3 &=& R_3 \star 1-\frac{1}{2}\star d\vec{\phi}\wedge d\vec{\phi}-\frac{1}{2}\epsilon_1\epsilon_2 e^{-\sqrt{3}\phi_1+\phi_2}\star \mathcal{F}_{(1)}\wedge \mathcal{F}_{(1)}\nn \\
& &-\frac{1}{2}\epsilon_1e^{-\sqrt{3}\phi_1-\phi_2}\star \mathcal{F}^1_{(2)}\wedge \mathcal{F}^1_{(2)}-\frac{1}{2}\epsilon_2e^{-2\phi_2}\star \mathcal{F}^2_{(2)}\wedge \mathcal{F}^2_{(2)},
\eea
where
\begin{align}
\mathcal{F}_{(1)}=&d\chi_1, &
\mathcal{F}_{(2)}^1=&d\mathcal{A}_{(1)}^1+\mathcal{A}_{(1)}^2\wedge d\chi_1, &
\mathcal{F}_{(2)}^2=&d\mathcal{A}^2_{(1)}, &
\end{align}
are the field strengths for $\chi_1$, $\mathcal{A}_{(1)}^1$, and $\mathcal{A}_{(1)}^2$ respectively. Adding the Lagrange multiplier terms
\be
-\chi_2 d(\mathcal{F}_{(2)}^1-\mathcal{A}_{(1)}^2\wedge d\chi_1) -\chi_3d\mathcal{F}_{(2)}^2,
\ee
and eliminating $\mathcal{F}_{(2)}^1$ and $\mathcal{F}_{(2)}^2$ we obtain the duality relations
\begin{align}
\epsilon_1 e^{-\sqrt{3} \phi_1 - \phi_2} \star \mathcal{F}_{(2)}^1 &=d\chi_2, &
\epsilon_2e^{-2 \phi_2} \star \mathcal{F}_{(2)}^2 &= d\chi_3-\chi_1d\chi_2.
\end{align}
In terms of the dualized variables the reduced three-dimensional Lagrangian becomes
\bea
\mathcal{L} &=& R \star 1-\frac{1}{2} \star d\vec{\phi}\wedge d\vec{\phi}-\frac{1}{2}\epsilon_1\epsilon_2e^{-\sqrt{3}\phi_1+\phi_2} \star d\chi_1\wedge d\chi_1
-\frac{1}{2}\epsilon_2 e^{\sqrt{3}\phi_1+\phi_2} \star d\chi_2\wedge d\chi_2
\nn \\
& &
-\frac{1}{2}\epsilon_1 e^{2\phi_2} \star (d\chi_3 - \chi_1 d \chi_2)\wedge (d\chi_3 - \chi_1 d \chi_2).
\label{3dLagF}
\eea

To obtain Lagrangian \eqref{3dLagF} from a coset construction we choose the basis for the fundamental representation of SL(3) where the Cartan-Weyl generators take the form,
\begin{align}
h_1 &=
\frac{1}{\sqrt{3}}\left(
\begin{array}{ccc}
 1 & 0 & 0 \\
 0 & -2 & 0 \\
 0 & 0 & 1
\end{array}
\right), &
h_2 &=
\left(
\begin{array}{ccc}
 1 & 0 & 0 \\
 0 & 0 & 0 \\
 0 & 0 & -1
\end{array}
\right),&
e_1 &=
\left(
\begin{array}{ccc}
 0 & 0 & 0 \\
 0 & 0 & 1 \\
 0 & 0 & 0
\end{array}
\right),&
e_2 &=
\left(
\begin{array}{ccc}
 0 & 1 & 0 \\
 0 & 0 & 0 \\
 0 & 0 & 0
\end{array}
\right), &
\label{basis}
\end{align}
and $e_3 = [e_2, e_1]$. The lowering generators are simply $f_i = e_i^T$. In this basis the positive roots are
\begin{align}
\alpha_1 &= (-\sqrt{3},1),&
\alpha_2 &= (\sqrt{3},1),&
\alpha_3 &= \alpha_1 + \alpha_2 = (0,2),&
\end{align}
and the negative roots are $- \alpha_1, -\alpha_2, - \alpha_3$.

We are interested in dimensional reduction of five-dimensional vacuum gravity over one timelike and one spacelike Killing direction. Since a timelike direction is involved, the standard Chevalley involution that expresses the symmetry between positive and negative roots does not define the coset of interest. The pertinent involution is,
\begin{align}
\tau(h_1) &= - h_1, & \tau(h_2) &= - h_2, &
\tau(e_1) &= - \epsilon_1 \epsilon_2  f_1, &
\tau(e_2) &= - \epsilon_2   f_2, &
\tau(e_3) &= - \epsilon_1 f_3, &
\label{involution}
\end{align}
where $\epsilon_{1,2} = \pm 1$.
 When $\epsilon_1 = \epsilon_2 = +1$ we get back the Chevalley involution.
The involution \eqref{involution} defines the generalized transposition
\be
x^\sharp = - \tau(x), \quad \forall  \quad x \in \mathfrak{sl}(3,\RR),
\ee
that can be implemented as matrix multiplication via
\be
x^\sharp = \eta x^T \eta, \qquad \mbox{where} \qquad \eta = \mathrm{diag}(1, \epsilon_2, \epsilon_1).
\label{gentranspose}
\ee
We note that $\eta^T = \eta^{-1} = \eta$. The Lie algebra generators that are invariant under the involution are
\begin{align}
k_1 &= e_1 - e_1^\sharp, &  k_2  &= e_2 - e_2^\sharp, &  k_3 &= e_3 - e_3^\sharp.
\label{kgenerators}
\end{align}
For the case $\epsilon_1 = -1, \epsilon_2 = + 1$ or $\epsilon_1 = + 1, \epsilon_2 = -1$  these generators form an $\mathfrak{so}(2,1)$ Lie algebra.

The three-dimensional scalar Lagrangian \eqref{3dLagF} can be parameterized by the SL(3,\,$\RR$)/SO(2,1) coset representative
\be
\mathcal{V}= e^{\frac{1}{2}\phi_1 h_1} e^{\frac{1}{2}\phi_2 h_2} e^{\chi_1e_{1}}e^{\chi_2e_{2}}e^{\chi_3e_{3}}.
\ee
From the coset representative we construct $M =   \cV^{\sharp} \cV.$ The three-dimensional Lagrangian can then be written as
\be
\cL'_3 = R\star 1 - \frac{1}{4} \mathrm{tr}(\star (M^{-1} d M) \wedge (M^{-1} d M)).
\ee
This form of the Lagrangian makes it manifestly invariant under SL(3,\,$\RR$).

\subsection{SU(2,\,1)/(SL(2,\,$\RR$) $\times$ U(1))}
\label{appendixA2}

Let us start by performing timelike Kaluza-Klein reduction from four to three dimensions of the four-dimensional Einstein-Maxwell theory. In our conventions the Lagrangian is
\be
\cL = R \star 1 - 2 \star F \wedge  F,
\ee
where $F = d A$. We reduce it to three-dimensions using the ansatz
\bea
ds^2_4 &=& - e^{-\phi}(dt + \omega)^2 + e^{\phi} ds^2_3, \label{metric4dEM} \\
A &=& \chi_e dt  + \tilde A \label{vector4dEM}.
\eea
All quantities on the right hand sides of equations \eqref{metric4dEM} and \eqref{vector4dEM} are independent of the time coordinate $t$. The reduced three-dimensional Lagrangian takes the form
\be
\cL_3 = R \star 1 - \frac{1}{2} \star d \phi \wedge d \phi + \frac{1}{2} e^{-2 \phi} \star \cF \wedge \cF -  2 e^{-\phi} \star \tilde F \wedge \tilde F + 2 e^\phi \star d\chi_e \wedge d \chi_e,
\ee
where
\be
\cF = d \omega, \qquad \qquad \tilde F = d \tilde A - d\chi_e \wedge \omega.
\ee

Adding the Lagrange multiplier terms
\be
- 4 d \chi_m \wedge \tilde F - (2 \chi_m d \chi_e - 2 \chi_e d \chi_m + \sqrt{2} d\psi) \wedge \cF,
\label{lag-mult}
\ee
and eliminating $\tilde F$ and $\cF$ we obtain the duality relations
\bea
\tilde F &=& - e^\phi \star d \chi_m, \\
\cF &=& e^{2\phi}\star(2\chi_m d \chi_e - 2 \chi_e d \chi_m + \sqrt{2} d\psi).
\eea
The dualized Lagrangian then takes the form
\bea
\cL_3'  &=& R \star 1 - \frac{1}{2} \star d \phi \wedge d \phi + 2 e^\phi (\star d\chi_e \wedge d \chi_e + \star d\chi_m \wedge d \chi_m) \nonumber \\
 & & - e^{2\phi} \star (d \psi + \sqrt{2} \chi_m d \chi_e - \sqrt{2} \chi_e d \chi_m) \wedge (d \psi + \sqrt{2} \chi_m d \chi_e - \sqrt{2} \chi_e d \chi_m).
\label{3dLagSU}
\eea
The Lagrange multiplier terms \eqref{lag-mult} are chosen in such a way that in the three-dimensional Lagrangian \eqref{3dLagSU} the electric and magnetic scalars $\chi_e$ and $\chi_m$ appear in a
symmetrical manner. In equation \eqref{3dLagSU} there are some sign changes compared to the standard spacelike reduction: the three-dimensional Lagrangian for that case can be obtained by a ``Wick rotation'' of the Maxwell scalars
\be
\chi_e  \to  - i \chi_e, \qquad \chi_m  \to i \chi_m, \qquad \psi \to - \psi.
\label{analyticC}
\ee
p
The scalar part of the three-dimensional Lagrangian  \eqref{3dLagSU} can be identified with the coset
$\mathrm{SU}(2,1)/(\mathrm{SL}(2,\RR) \times \mathrm{U}(1))$. We describe this construction in the rest of this appendix.
For the case of the spacelike reduction the corresponding coset is
\be
\mathrm{SU}(2,1)/(\mathrm{SU}(2) \times \mathrm{U}(1)).
\ee
Naturally, the change in the denominator group has its origin in different signs for the kinetic terms in
\eqref{3dLagSU} corresponding to the timelike or spacelike reduction.

In order to describe the coset construction, let us start by recalling some basic properties of the group SU(2,1).
In our conventions the group SU(2,1) is defined by the set of unit determinant $(3 \times 3)$ \emph{complex} matrices $g$ that preserve a metric $\kappa$ of
signature $(+, +, -)$:
\be
\mathrm{SU}(2,1) = \left\{ g \in \mathrm{SL}(3, \mathbb{C}): g^\dagger \kappa g = \kappa \right\} \qquad \mbox{with} \qquad \kappa =
\left(
\begin{array}{ccc}
0 & 0 & -1 \\
0 & 1 & 0 \\
-1 & 0 & 0
\end{array}
\right).
\label{kappa}
\ee
The associated Lie algebra is denoted as $\mathfrak{su}(2,1)$. The  $\mathfrak{su}(2,1)$ Lie algebra is a non-split real form of $\mathfrak{sl}(3, \mathbb{C})$.
 In the basis \eqref{basis}\footnote{Recall that the real span of the $\mathfrak{sl}(3, \mathbb{C})$ generators in the Cartan-Weyl basis gives the $\mathfrak{sl}(3, \mathbb{R})$ Lie algebra --
the split real form of $\mathfrak{sl}(3, \mathbb{C})$.} it
 is described by the real span of the following linear combinations of the $\mathfrak{sl}(3, \mathbb{C})$ generators
\be
 \{i \sqrt{3} h_1, h_2, e_1 + e_2, f_1 + f_2, i (e_2 - e_1), i (f_2-f_1), i e_3, i f_3\}.
\ee
It can be readily checked using the matrix representation given above that these linear combinations of generators satisfy $x^\dagger \kappa  + \kappa x  = 0$.
The generators $ \{i \sqrt{3} h_1, h_2 \}$ belong to the Cartan subalgebra of $\mathfrak{su}(2,1)$, $\{ e_1 + e_2, i (e_2 - e_1),i e_3 \}$ are
the positive generators while $\{ f_1 + f_2, i (f_2-f_1), i f_3\}$ are the negative generators.
The two subalgebras that play important role in our analysis are  $(i)$ the maximally compact subalgebra
\be
\mathfrak{su}(2)  \oplus \mathfrak{u}(1) = \{x \in \mathfrak{su}(2,1): x^\dagger = -x\},
\ee
that defines the $\mathrm{SU}(2,1)/(\mathrm{SU}(2) \times \mathrm{U}(1))$ coset,
and $(ii)$ the maximally non-compact subalgebra
\be
\mathfrak{sl}(2,\RR) \oplus \mathfrak{u}(1) = \{x \in \mathfrak{su}(2,1): x^\dagger = - \eta x \eta^{-1} \},
\ee
where $\eta = \mbox{diag} \{ 1,-1,1 \}$ that defines the $\mathrm{SU}(2,1)/(\mathrm{SL}(2,\RR) \times \mathrm{U}(1))$ coset. Explicit linear combinations of generators
that make these subalgebras manifest can be found in \cite{Houart:2009ed}. We define
generalized transposition as
\be
x^\sharp :=  \eta x^\dagger \eta^{-1} \quad \forall \quad x \in \mathfrak{su}(2,1).
\label{SUgentranspose}
\ee

The three-dimensional scalar Lagrangian in equation \eqref{3dLagSU} can be  parameterized by the coset representative (see e.g.~reference \cite{Houart:2009ed})
\be
\cV = \exp\left[ \frac{1}{2} \phi h_2 \right] \cdot \exp\left[ \sqrt{2} \chi_e (e_1 + e_2) + \sqrt{2}\chi_m (i (e_2 - e_1)) + \sqrt{2} \psi (i e_3) \right].
\ee
From the coset representative we construct
\be
M =   \cV^{\sharp} \cV.
\ee
The three-dimensional Lagrangian \eqref{3dLagSU} can now be written as
\be
\cL'_3 = R\star 1 - \frac{1}{4} \mathrm{tr}(\star (M^{-1} d M) \wedge (M^{-1} d M)).
\ee
This form of the Lagrangian makes it manifestly invariant under SU(2,1) with $M \to M' = g^\sharp M g$, where $g$ is any SU(2,1) matrix.

\end{document}